\documentclass[conference]{IEEEtran}
\IEEEoverridecommandlockouts
\usepackage{amsmath,amssymb,amsfonts}
\usepackage{algorithmic}
\usepackage{graphicx}
\usepackage{textcomp}
\usepackage{xcolor}
\def\BibTeX{{\rm B\kern-.05em{\sc i\kern-.025em b}\kern-.08em
    T\kern-.1667em\lower.7ex\hbox{E}\kern-.125emX}}

\usepackage{tabularx, booktabs}
\usepackage{mathtools, nccmath, amssymb}
\usepackage{hyperref}
\usepackage{tabu}
\usepackage{subcaption}

\newcommand{\sstitle}[1]{\smallskip\noindent\textbf{#1.\/}}

\newcommand{\ie}{i.\,e.,\ }

\begin{document}

\title{Efficient ECG-based Atrial Fibrillation Detection via Parameterised Hypercomplex Neural Networks\\
\thanks{This research was funded by the Federal Ministry of Education and Research (BMBF), Germany under the project LeibnizKILabor with grant No.\,01DD20003.}
}

\author{\IEEEauthorblockN{Leonie Basso}
\IEEEauthorblockA{\textit{L3S Research Center} \\
\textit{Leibniz Universit\"at Hannover}\\
Hannover, Germany \\
basso@l3s.de}
\and
\IEEEauthorblockN{Zhao Ren}
\IEEEauthorblockA{\textit{L3S Research Center} \\
\textit{Leibniz Universit\"at Hannover}\\
Hannover, Germany \\
zren@l3s.de}
\and
\IEEEauthorblockN{Wolfgang Nejdl}
\IEEEauthorblockA{\textit{L3S Research Center} \\
\textit{Leibniz Universit\"at Hannover}\\
Hannover, Germany \\
nejdl@l3s.de}
}

\maketitle

\begin{abstract}
Atrial fibrillation (AF) is the most common cardiac arrhythmia and associated with a high risk for serious conditions like stroke. 
The use of wearable devices embedded with automatic and timely AF assessment from electrocardiograms (ECGs) has shown to be promising in preventing life-threatening situations.
Although deep neural networks have demonstrated superiority in model performance, their use on wearable devices is limited by the trade-off between model performance and complexity.
In this work, we propose to use lightweight convolutional neural networks (CNNs) with parameterised hypercomplex (PH) layers for AF detection based on ECGs.
The proposed approach trains small-scale CNNs, thus overcoming
the limited computing resources on wearable devices.
We show comparable performance to corresponding real-valued CNNs on two publicly available ECG datasets using significantly fewer model parameters.
PH models are more flexible than other hypercomplex neural networks and can operate on any number of input ECG leads.
\end{abstract}

\begin{IEEEkeywords}
Atrial fibrillation, ECG analysis, hypercomplex domain, lightweight neural networks
\end{IEEEkeywords}

\section{Introduction}
\label{sec:intro}
Atrial fibrillation (AF), characterised by an irregular beating of the atrial chambers of the heart, is the most common serious abnormal heart rhythm and a risk factor for strokes~\cite{Lippi2021}.
A timely diagnosis of AF is crucial for the patient's health. 
The most reliable way to test for cardiac arrhythmia is to analyse the electrical activity of a heart as electrocardiograms (ECGs)~\cite{Rizwan2021}.
However, diagnosis of AF from ECGs requires well-trained medical professionals, and AF often remains unrecognised during conventional short-time monitoring due to its episodic nature in early stages~\cite{Rizwan2021}.
In the era of Internet-of-Medical-Things, wearable devices such as Holter monitors and smartwatches are enabling the diagnosis of cardiac diseases in daily life based on long-interval ECG signals~\cite{Rizwan2021,sajeev2019wearable}.
A manual diagnosis from long-term ECG signals by medical professionals is time-consuming~\cite{Rizwan2021}, which results in a bottleneck for self-care of AF. Therefore, automated analysis of long-interval ECG signals is essential for AF detection.
Recently, deep neural networks (DNNs) have achieved success in ECG signal analysis~\cite{Hong2020}, such as arrhythmia detection~\cite{Hannun2019} and AF classification~\cite{Andersen2019}.
However, the complexity of large DNNs limits their deployment on wearable devices~\cite{Hong2020}, which motivates the need for small and efficient models.

A variety of model compression techniques have been proposed.
For example, pruning aims to discard unnecessary network connections and quantisation targets to represent model weights with less bits~\cite{Han2015}. 
Knowledge distillation builds a shallow student model trained with a pre-trained deep teacher model~\cite{Hinton2015, polino2018model}.
In contrast to real-valued neural works used in the above approaches, quaternion neural networks (QNNs) build lightweight neural networks by inherently changing the construction of layer weights.
The input elements are processed as entities of one real and three imaginary components based on 4-dimensional hypercomplex quaternion numbers~\cite{Parcollet2020}.
QNNs only need \(1/4\) learnable parameters compared to corresponding real-valued neural networks and can capture internal latent relations between input channels~\cite{Parcollet2020}. 
Nevertheless, their dimensionality is limited because operations in hypercomplex space are only predefined in limited dimensions, such as 4D, 8D, and 16D.

Parameterised hypercomplex (PH) neural networks~\cite{Zhang2021,Grassucci2021} were proposed to overcome the dimensionality restrictions of QNNs.
They can operate in any input dimension, allowing a flexible and domain-specific processing of multidimensional inputs.
PH layers use Kronecker product properties to construct weight matrices that reduce the number of learnable parameters to \(1/n\) compared to corresponding real-valued neural networks based on the hyperparameter \(n\).
Similar to QNNs, PH models are also able to learn internal 
relations between channels.
Specifically, PH multiplication (PHM)~\cite{Zhang2021} has been proposed to replace real-valued matrix multiplication in fully-connected (FC) layers.
PH convolutional (PHC) layers~\cite{Grassucci2021} extend PHM to convolution and facilitate the development of efficient deep convolutional neural networks (CNNs). 
Therefore, PH models can help apply deep learning to AF detection on wearable devices by using significantly fewer parameters than real-valued models.

To the best of the authors' knowledge, hypercomplex DNNs have rarely been applied to ECG signals. 
The study in~\cite{cruces2018quaternion} used QNNs to model cardiac velocity patterns from ECG-related vectorcardiogram signals, which record spatio-temporal information of cardiac electrical activity.
The 4D quaternions are suitable to model rotations in a 3D space, however, not applicable to ECG signals with different numbers of leads.
In this work, we propose lightweight PH-CNNs with PHM and PHC layers
for AF detection and abnormality classification from ECG signals. 
Our contributions are manifold.
(1) PH-CNNs can significantly reduce the number of model parameters and achieve comparable performance to corresponding real-valued models, enabling applications of effective AF detection on wearable devices. 
(2) PH-CNNs are flexible in input dimensions and can effectively process ECG signals with different numbers of leads.
(3) Learning inter-channel relations between the different ECG leads with PH-CNNs is promising to improve the model performance.
We conduct experiments to demonstrate the performance and efficiency of the approach with several CNN architectures
and channel dimensionalities on two publicly available
datasets.

\section{Methodology}
\label{sec:methods}
\begin{figure}[t]
  \centering
  \includegraphics[width=\linewidth]{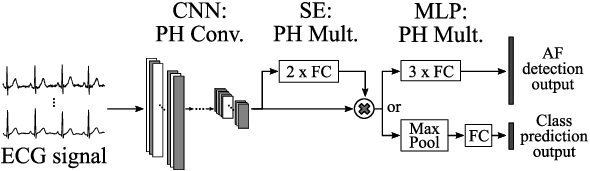}
\caption{Proposed PH-CNN architecture, including three modules: 
(1)~a~CNN, (2) a squeeze-and excitation (SE) attention, and (3) a multilayer perceptron (MLP) classifier.
Compared to real-valued DNNs, parameterised hypercomplex (PH) convolution and multiplication replace real-valued convolutional and fully-connected (FC) layers, respectively.
We construct separate models for two tasks: (a) AF detection, where every sampling point of the input ECG signal gets classified as AF/non-AF, and (b) global abnormality classification, where the output is a vector of class probabilities.}
\label{fig:dcnn}
\end{figure}
We propose a PH-CNN architecture that takes raw ECG signals as input and
consists of three modules (see \figurename~\ref{fig:dcnn}):
(1) A CNN with PHC layers extracts features, (2) a squeeze-and-excitation (SE)~\cite{Hu2018} attention mechanism with PHM improves channel interdependencies,
and (3) a multilayer perceptron (MLP) classifier with PHM produces predictions.
We construct separate models for two objectives: (1) \emph{AF detection} and (2) \emph{rhythm/morphology abnormality classification}.
AF detection aims to detect the on- and offsets of AF episodes by predicting AF or non-AF for each sampling point, which helps in developing personalised treatments~\cite{de2020temporal}. The classification of abnormalities, including AF, is performed at the global level for each ECG recording.

\subsection{Parameterised Hypercomplex Neural Network}
\label{sec:phc}
A real-valued FC layer has a weight matrix \(\mathbf{W}_{\text{FC}}\!\in\!\mathbb{R}^{d_{\text{out}} \times d_{\text{in}}}\) for in- and output dimensions \(d_{\text{in}}\) and \(d_{\text{out}}\). Thus, the parameter size of a real-valued FC layer is \(\mathcal{O}(d_{\text{out}}\cdot {d_{\text{in}}})\).
The parameter size of a real-valued 1D convolutional layer is \mbox{\(\mathcal{O}(d'_{\text{out}} \cdot  d'_{\text{in}} \cdot k)\)} for a weight matrix \(\mathbf{W}_{\text{Conv}}\!\in\!\mathbb{R}^{d'_{\text{out}} \times d'_{\text{in}} \times k}\), in- and output dimensions \(d'_{\text{in}}\) and \(d'_{\text{out}}\), and the size \(k\) of a 1D kernel. 

\begin{figure}[t]
  \centering
  \includegraphics[width=\linewidth]{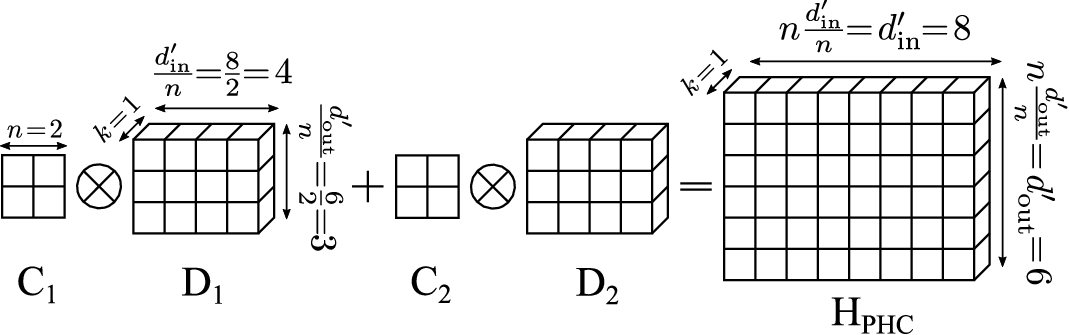}
\caption{Example of a weight matrix \(\mathbf{H}_{\text{PHC}}\) in a parameterised hypercomplex convolutional (PHC) layer~\cite{Grassucci2021} 
when \(n\!=\!2\).
\(\mathbf{H}_{\text{PHC}}\) is built by the sum of the Kronecker products of matrices \(\mathbf{C}_1,\mathbf{C}_2\!\in\!\mathbb{R}^{n\times n}\) and \(\mathbf{D}_1, \mathbf{D}_2\!\in\!\mathbb{R}^{(d'_{\text{out}}/n) \times (d'_{\text{in}}/n) \times k}\). The row or column size of a Kronecker product is the multiplication of the row or column size of the input matrices, respectively, therefore \(\mathbf{H}_{\text{PHC}}\!\in\!\mathbb{R}^{d'_{\text{out}} \times d'_{\text{in}} \times k}\). In this example, the input channel size \(d_{in}\!=\!8\), output size \(d_{out}\!=\!6\), and kernel size \(k\!=\!1\). 
(Adapted from~\cite{Zhang2021})} 
\label{fig:phc_layers}
\end{figure}

PHM~\cite{Zhang2021} and PHC~\cite{Grassucci2021} layers generalise hypercomplex operations and 
can operate in arbitrary domains \(n\)D, where the dimensionality \(n\!\in\!\mathbb{N}\) can be chosen depending on the input dimensionality or adapted as a hyperparameter. The input elements are processed as hypercomplex numbers of dimension \(n\). Due to weight sharing over the \(n\) dimensions, PH models are able to capture internal relations between channels.

In a PHM layer~\cite{Zhang2021}, the multiplication of an input \(x\!\in\!\mathbb{R}^{d_{\text{in}}}\) or \(x\!\in\!\mathbb{R}^{d_{\text{in}} \times t}\) with a weight matrix \(\mathbf{H}_{\text{PHM}}\!\in\!\mathbb{R}^{d_{\text{out}} \times d_{\text{in}}}\) to produce an output \(y\!\in\!\mathbb{R}^{d_{\text{out}}}\) or \(y\!\in\!\mathbb{R}^{d_{\text{out}} \times t}\) is defined as
\begin{equation}
    y = \text{PHM}(x) =\mathbf{H}_{\text{PHM}} \cdot x + \mathbf{b}_{\text{FC}}.
\end{equation}
\(\mathbf{H}_{\text{PHM}}\) is built by the sum of \(n\) Kronecker products 
of learnable matrices \(\mathbf{A}_i\!\in\!\mathbb{R}^{n \times n}\) 
and \(\mathbf{B}_i\!\in\!\mathbb{R}^{ (d_{\text{out}}/{n}) \times (d_\text{in}/{n})}\) for \mbox{\(i\!=\!1,\dots,n\)}, 
\begin{equation}
    \mathbf{H}_{\text{PHM}} = \sum_{i=1}^{n} \mathbf{A}_i \otimes \mathbf{B}_i.
\end{equation} 
The hypercomplex multiplication of the weight parameters \(\mathbf{B}\) and the input \(x\) relies on interactions between imaginary units, which are modelled by the algebra rules in \(\mathbf{A}_i\). These define arithmetic operations, such as multiplication, in the \(n\)D hyperspace and determine the arrangement of \(\mathbf{B}_i\) in the final weight matrix \(\mathbf{H}_{\text{PHM}}\).
PH layers learn the algebra rules as matrices \(\mathbf{A}_i\)  directly from the data.
Thus, they don't rely
on predefined rules, which exist, for example, for the complex (\ie 2D) and quaternion (\ie 4D) domains but not for 6D.

In a PHC layer~\cite{Grassucci2021}, the convolution of an input \(x'\!\in\!\mathbb{R}^{d'_{\text{in}} \times t_{\text{in}}}\) with a weight matrix \(\mathbf{H}_{\text{PHC}}\) to produce an output \(y'\!\in\!\mathbb{R}^{d'_{\text{out}} \times t_{\text{out}}}\) is defined as
\begin{equation}
    y' = \text{PHC}(x') =\mathbf{H}_{\text{PHC}} \ast x' + \mathbf{b}_{\text{Conv}}.
\end{equation}
Analogous to \(\mathbf{H}_{\text{PHM}}\), 
\(\mathbf{H}_{\text{PHC}}\!\in\!\mathbb{R}^{d'_{\text{out}} \times d'_{\text{in}} \times k}\) is built by the sum of \(n\) Kronecker products of learnable matrices \(\mathbf{C}_i\!\in\!\mathbb{R}^{n \times n}\) and \(\mathbf{D}_i\!\in\!\mathbb{R}^{(d'_{\text{out}}/n) \times (d'_{\text{in}}/n) \times k}\) for a 1D kernel with size \(k\) (see Fig~\ref{fig:phc_layers}). 
\(\mathbf{D}_i\) represents the \(i\)-th collection of convolution filters that are organised after the algebra rules \(\mathbf{C}_i\),
\begin{equation}
    \mathbf{H}_{\text{PHC}} = \sum_{i=1}^{n} \mathbf{C}_i \otimes \mathbf{D}_i.
\end{equation}
The parameter size of a PHM layer can be approximated to \mbox{\(\mathcal{O}(d_{\text{out}}\cdot d_{\text{in}}/n)\)} and of a PHC layer to \(\mathcal{O}(d'_{\text{out}}\cdot d'_{\text{in}}\cdot k/n)\), if \mbox{\(d_{\text{out}}\cdot d_{\text{in}} \gtrapprox n^4\)} holds. This assumption is mild for real-world problems, where usually large numbers of filters are employed.
This leads to a parameter reduction to approximately \(1/n\) compared to a standard real-valued FC or convolutional layer, respectively.
Note that the dimensions \(d_\text{in}\), \(d_\text{out}\), \(d'_\text{in}\) and \(d'_\text{out}\) have to be divisible by \(n\).

\subsection{Model architectures}

\begin{figure}[t]
\begin{subfigure}[t]{0.534\linewidth}
    \centering
    \includegraphics[width=\hsize]{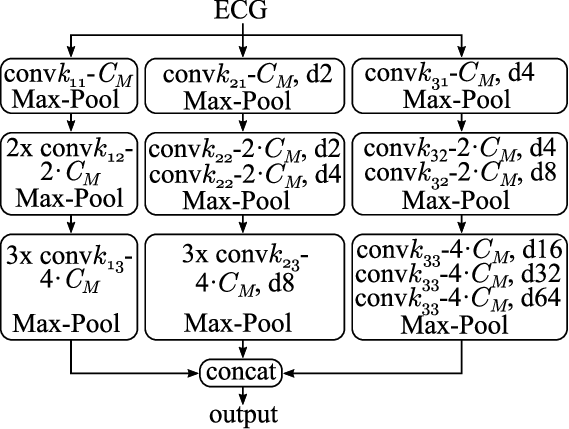}
    \caption{Multi-Scopic}
    \label{fig:cnns-ms}
\end{subfigure}
\begin{subfigure}[t]{0.277\linewidth}
    \centering
    \includegraphics[width=\hsize]{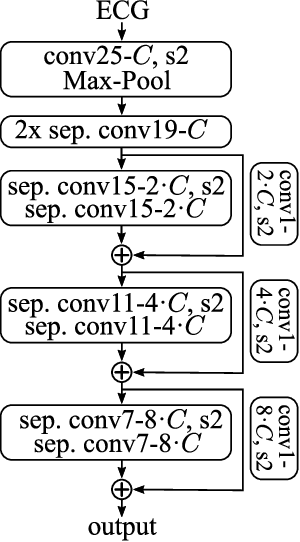}
    \caption{ResNet}
    \label{fig:cnns-res}
\end{subfigure}
\begin{subfigure}[t]{0.165\linewidth}
    \centering
    \includegraphics[width=\hsize]{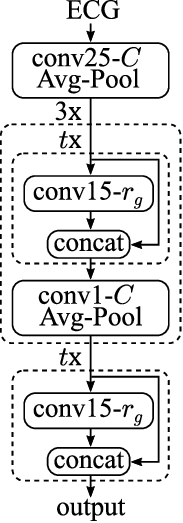}
    \caption{Dense}
    \label{fig:cnns-dense}
\end{subfigure}
\caption{Configurations of the three CNN backbones. Convolutional layers are denoted as ``conv(kernel size\,-\,number of channels)'' with ``s(stride size)'' and ``d(dilation rate)'' (default 1). The output is then fed into an attention module (see \figurename~\ref{fig:dcnn}). The selected parameter values depend on the input dimensionality: for 2-channel input
\(C_M{=}16\), \(C{=}64\), \(t{=}4\), growth rate \(r_g{=}16\), \(k{=}[[3,3,3],[5,5,3],[9,7,5]]\), 
and for 12-channel input \(C_M{=}24\), \(C{=}72\), \(t{=}6\), \(r_g{=}12\), \(k{=}[3{\times}[11,7,5]]\).} 
\label{fig:cnns}
\end{figure}

We select three lightweight CNN backbones: Multi-Scopic CNN~\cite{Cai2020}, ResNet~\cite{He2016}, and DenseNet~\cite{Huang2017}.
The Multi-Scopic CNN has been used as a default model for an AF detection benchmark~\cite{torch_ecg_paper2} and both ResNets and DenseNets have been successfully applied for arrhythmia classification from ECGs~\cite{Hannun2019,torch_ecg_paper2}.
The CNN model configurations 
are shown in \figurename~\ref{fig:cnns}.
The Multi-Scopic CNN consists of three parallel CNN branches that use different dilation rates to combine information from different receptive fields.
The DenseNet architecture consists of several dense blocks, which concatenate output feature maps of a layer with input feature maps and are connected by transition blocks. 
In ResNets, the output of a residual block is added to the input that is upsampled in a shortcut connection.
To reduce model parameters, the residual blocks here use separable convolutions, which consist of a depth-wise and a point-wise operation.
In depth-wise convolutions, the input channels are processed separately with \(d'_{in}\!=\!1\). Therefore, PHC is not used for depth-wise convolution, but for the \(1\!\times\!1\) point-wise convolutions in separable convolutions and in the shortcut connection of each residual block.

Each CNN module is followed by an SE block~\cite{Hu2018}, which performs feature recalibration by multiplying input feature maps with channel weights learnt by two FC layers.
The MLP classifier for AF detection consists of three FC layers and a linear interpolation to recover the sequence length of the inputs.
For abnormality classification, a probability for each class is produced by max pooling and an FC layer (see \figurename~\ref{fig:dcnn}). 

\section{Experiments and Results}
\label{sec:exp_results}

\subsection{Datasets}
\label{sec:data}
For this study, we use two datasets with different sources, number of recorded ECG leads, and tasks.

\begin{table}[t]
    \centering
    \caption{Distribution of classes in the training, test and validation sets for segments from the CPSC 2021 and CPSC 2018 datasets.}   
    \vspace{-5pt}
    \scalebox{0.89}{
    \begingroup
    \setlength{\tabcolsep}{2pt} 
    \hspace*{-0.25\leftmargin}
    \begin{tabu}[numbersep=0pt,resetmargins=true]{@{} l|cccc|cccccccc}
    \toprule
    \rowfont{\small}
    &\multicolumn{4}{c|}{CPSC 2021} & \multicolumn{7}{c}{CPSC 2018}\\
    \rowfont{\scriptsize} 
    \# & non-AF & pers. AF & PAF & \multicolumn{1}{c|}{\(\sum\)} & NSR & AF & I-AVB & LBBB & RBBB & PAC & \(\sum\) \\
    \midrule
    \rowfont{\small}
         Train& 35,643 & 12,442 & 6,169 & 54,254 & 374 & 508 & 290 & 96 & 729 & 232 & 2229 \\
    \rowfont{\small}
         Val.& 23,146 & 14,382 & 1,947 & 39,475 & 261 & 358 & 225 & 66 & 559 & 192 & 1661 \\
    \rowfont{\small}
         Test& 17,244 & 11,317 & 3,153 & 31,714 & 283 & 354 & 206 & 73 & 566 & 190 & 1672 \\
    \bottomrule
    \end{tabu}
    \endgroup
    }
    \label{tab:class_dist}
\end{table}

\sstitle{CPSC 2021}
We use 2-lead data from the CPSC 2021 dataset~\cite{Wang2021}
for the detection of paroxysmal 
atrial fibrillation (PAF) events, which are intermittent AF episodes that end within less than a week. 
The data include signals from lead~I and lead~II of long-term dynamic ECGs recorded with 
wearable Holter monitors at 200\,Hz.
The publicly available training set consists of 1,436 records 
from 23 PAF, 24 persistent AF, and 53 non-AF patients.
Due to the unreleased test set, we perform a random subject-independent split in 40\,\% train, 30\,\% validation, and 30\,\% test sets. 
All classes appear in all subsets with approximately the same distribution as in the entire dataset (see Table~\ref{tab:class_dist}).
Apart from AF detection, the global rhythm is derived from the AF label of each sampling point in order to perform a three-class classification:
\emph{non-AF} (no AF rhythm present),
\emph{PAF} (both AF and non-AF episodes), and \emph{persistent AF} (continuous AF).
During preprocessing, z-score normalisation and a bandpass filter with cutoff frequencies of 0.5 and 45\,Hz were applied, following~\cite{Wen2022:cpsc2021-work}.
The preprocessed signals were then sliced into 30\,s segments with 15\,s overlap. 

\sstitle{CPSC 2018}
We use 12-lead ECG signals from the CPSC 2018 dataset~\cite{CPSC2018} as provided in the PhysioNet/Computing in Cardiology Challenge (CinC)
2020~\cite{CinC2020,PhysioNet}. 
The dataset includes 5,271 records that are collected from in-hospital patients at 500 Hz. 
Again, we split the data subject-independently into 40\,\% train, 30\,\% validation and 30\,\% test sets considering class balance (see Table~\ref{tab:class_dist}).
A multi-label classification task is performed with the following rhythm/morphology classes:
normal sinus rhythm (NSR), AF, first-degree atrioventricular block (I-AVB), left bundle branch block (LBBB), right bundle branch block (RBBB), and premature atrial contraction (PAC). 
The signals are z-score normalised
and a 10\,s segment is extracted from the middle of each record. 

\subsection{Experimental Setup}
For CPSC 2021, we run experiments with \(n\!=\!2\) in the PH models, matching the number of ECG leads, and also with \(n\!=\!4\), where the input signals are zero-padded to 4 channels.
For CPSC 2018, we test \(n\!=\!2,4,6,12\) 
to investigate a trade-off between accuracy and model complexity.
The value of \(n\) is the same for all layers except for the final MLP layers, where \(n\!=\!1\) because 
\(d_{\text{out}}\) must be divisible by \(n\).

Direct comparison with previous studies is difficult due to different data sets and splits.
We compare the performance of our framework with two popular deep ResNets for arrhythmia classification~\cite{Hannun2019,Ribeiro2020:ecg}, and a Multi-Scopic CNN with SE attention and an additional bidirectional long short-term memory (LSTM)~\cite{Wen2022:cpsc2021-work},
following the benchmarks proposed in~\cite{torch_ecg_paper2}.

\subsection{Evaluation Metrics}
We evaluate the classification performance based on the unweighted average recall (UAR) to account for class imbalance. 
For the classification of classes \(C\), the UAR is 
based on true-positive (TP) and false-negative (FN) samples,
\begin{equation}
    \text{UAR} = \frac{1}{|C|} \sum_{c\in C} \frac{\text{TP}_c}{\text{TP}_c+\text{FN}_c}.
\end{equation}
To assess the significance of the comparison between UARs of PH-CNNs and real-valued networks, we perform a one-tailed z-test on the test sets.
In addition, the challenge score \(S_{\text{AF}}\) of CPSC 2021 \cite{Wang2021} is used to evaluate the correct detection of AF events. 
\(S_{\text{AF}}\) evaluates the overall rhythm prediction as well as the detection of start and end points of an AF episode.

\subsection{Model Training}

The models are implemented using the \verb|torch_ecg| package~\cite{torch_ecg, torch_ecg_paper2}.
We generally follow the training setup proposed in their CPSC 2021 and CinC 2020 (for CPSC 2018) benchmark tests.
We use the Asymmetric Loss~\cite{Ridnik2021_AsymmetricLoss} to handle the class imbalance, a learning rate of 0.0001 and the AMSGrad variant of the AdamW optimiser~\cite{Reddi2018_amsgrad}. 
The best model is determined based on binary-cross-entropy (for AF detection) or global accuracy (for abnormality classification) on the validation set.
The batch size is set to 64. 
The code is available at: \href{https://github.com/leibniz-future-lab/HypercomplexECG}{https://github.com/leibniz-future-lab/Hypercomplex-ECG}. 

\subsection{Results and Discussion}
\label{sec:results}
\begin{table}[t]
\centering
\caption{Comparison between parameterised hypercomplex (PH) models with hyperparameter \(n\) and real-valued models (\(n\) is \mbox{`-')} for AF detection on the 2-lead ECG data in CPSC 2021. 
Results are reported for Unweighted Average Recall (UAR) and Score \(S_{\text{AF}}\) from CPSC 2021 on the validation and test sets. C.Par. = CNN Parameters, Multi-Sc. = Multi-Scopic.}
\begingroup
\setlength{\tabcolsep}{2.5pt} 
\begin{tabular}{lc
rr cccc
}
    \toprule
    Network &\(n\)& \#C.Par. & \#Params & \multicolumn{2}{c}{UAR in \% \(\uparrow\)} & \multicolumn{2}{c}{\(S_{\text{AF}}\) \(\uparrow\)} \\ 
    \cmidrule(lr){5-6} \cmidrule(lr){7-8}
    &&& & Val & Test & Val & Test \\  
    \midrule
    ResNet \cite{Hannun2019} & -- & 26,499k & 26,746k & \textbf{87.40} & 83.86 & 1.3719 & 1.3804 \\
    ResNet \cite{Ribeiro2020:ecg} & -- & 8,708k & 9,463k & 81.24 & 82.04 & 1.3389 & 1.3622 \\
    Multi-Sc.+LSTM\,\cite{Wen2022:cpsc2021-work} & -- & 138k & 2,850k & 84.16 & \textbf{87.14} & \textbf{1.3917} & \textbf{1.4258} \\
    \midrule
    Multi-Scopic &--& 138k & 213k & 81.25 & 83.51 & \textbf{1.8944} & 2.0486 \\
    PH-Multi-Scopic  &2& 71k & 112k & 82.06 & \textbf{86.06} & 1.6170 & 2.0560 \\ 
    PH-Multi-Scopic  &4& 38k & 61k & \textbf{88.79} & 82.62 & 1.6989 & \textbf{2.4488} \\
    \midrule[0.5pt]
    ResNet &--& 720k & 934k & 80.14 & 86.17 & 1.3392 & 1.4622 \\
    PH-ResNet  &2& 373k & 497k & 86.22 & 81.88 & 1.3721 & 1.4424 \\
    PH-ResNet  &4& 200k & 279k & \textbf{88.61} & \textbf{86.68} & \textbf{1.4198} & \textbf{1.4629} \\
    \midrule[0.5pt]
    DenseNet &--& 369k & 423k & \textbf{87.29} & \textbf{83.49} & \textbf{1.4065} & 1.3636 \\
    PH-DenseNet &2& 187k & 215k & 83.44 & 78.80 & 1.3523 & \textbf{1.3672} \\
    PH-DenseNet &4& 98k & 113k & 76.26 & 78.72 & 1.3490 & 1.3431 \\
    \bottomrule
\end{tabular}
\endgroup
\label{tab:results_2lead}
\end{table}

\sstitle{AF Detection} 
Using PH-Multi-Scopic as the CNN backbone yielded the highest score \(S_{\text{AF}}\) for detecting the start and end points of AF episodes on CPSC 2021 data (see Table~\ref{tab:results_2lead}). 
This value is also comparable to most challenge submissions~\cite{Wang2021}, however, a direct comparison is difficult due to different test sets. 
PH-Multi-Scopic (\(n{=}2\), \(p{<}0.001\) in one-tailed z-test)
and PH-ResNet (\(n{=}4\), \(p{<}0.05\))
models achieve 
a higher UAR than the corresponding real-valued networks. A possible explanation is the ability of PH layers to learn inter-channel relations due to weight sharing. 
These two models still achieve high UARs when the model is further compressed with \(n\!=\!4\). 
However, the UAR decreases for PH-DenseNet (\(n{=}2\), \(p{<}0.001\)), while \(S_{\text{AF}}\) is comparable. 
Both metrics are calculated from the same sequence output, but the UAR weights all classes equally
and doesn't evaluate the AF 
on-/offsets.
The PH-DenseNets generalise worse on the PAF class in the validation and test sets, possibly due to DenseNet's simpler structure and limited ability to learn useful features from this data.
The previously proposed models~\cite{Hannun2019,Ribeiro2020:ecg,Wen2022:cpsc2021-work}
achieve comparable UARs to the best proposed small-scale model, while using significantly more parameters.
They also have a notably lower \(S_{\text{AF}}\) than PH-Multi-Scopic and are therefore worse at detecting the start and end of AF episodes.

\sstitle{Abnormality Classification}
The classification performance of PH and real-valued models on the 12-lead CPSC 2018 data is comparable for all three CNN backbones (see Table~\ref{tab:results_12lead}).
The highest UAR overall is achieved by PH-Multi-Scopic models.
An increased \(n\) doesn't lead to a declining performance, therefore, we observe no large trade-off between increased parameter saving and classification performance in this setting. This could again be explained with the ability of PH models to capture the relations between different channels, especially for a larger \(n\) that covers all input channels.
Again, the previously proposed models~\cite{Hannun2019,Ribeiro2020:ecg,Wen2022:cpsc2021-work} don't exceed the classification performance of the small-scale models while using significantly more parameters (\(p{<}0.05\) for PH-Multi-Scopic \(n\!=\!12\) and ResNet~\cite{Hannun2019}).

\sstitle{Model compression}
A large parameter reduction can be observed using PHC layers in all CNNs, saving already approximately \mbox{-50\,\%} CNN parameters for \(n\!=\!2\).
The number of parameters is further reduced as \(n\) increases,
leading to a substantial model compression when \(n\!\!=\!\!12\) on 12-lead data, where the number of parameters in all three CNNs is reduced by more than 80\%.
Even though the depth-wise convolutions in ResNet couldn't be adapted to PHC, the number of parameters are notably reduced due to PH point-wise convolutions.
The CNN module accounts for the vast majority of the model size, so its compression has the greatest impact.
PHM layers also reduce the parameter size of the FC layers in the SE and MLP module to approximately \(1/n\), resulting in a large overall parameter reduction.
The value of \(n\) can be chosen flexibly during model creation based on the data dimensionality and desired model compression.

\begin{table}[t]
\centering
\caption{Comparison between parameterised hypercomplex (PH) models with hyperparameter \(n\) and real-valued models (\(n\) is \mbox{`-')} for abnormality classification on 12-lead ECG data from CPSC 2018. Results are reported for Unweighted Average Recall (UAR) on the validation and test sets.}
\begingroup
\setlength{\tabcolsep}{4pt} 
\begin{tabular}{lcr lr cc
}
    \toprule
    Network  & \(n\)& \multicolumn{2}{c}{\#CNN Params} & \#Params & \multicolumn{2}{c}{UAR in \% \(\uparrow\)} \\ 
    \cmidrule(lr){6-7}
    & & & & & Val & Test \\ 
    \midrule
    ResNet \cite{Hannun2019} & -- & 26,511k & & 26,514k & \textbf{92.43} & \textbf{90.34} \\
    ResNet \cite{Ribeiro2020:ecg} & -- & 8,719k & & 8,726k & 91.71 & 89.09 \\
    Multi-Sc.+LSTM\,\cite{Wen2022:cpsc2021-work} & -- & 430k & & 2,135k & \textbf{92.43} & 89.99 \\
    \midrule
    Multi-Scopic & --& 430k & & 473k & 90.24 & 90.68 \\
    PH-Multi-Scopic & 2 & 219k & (-49\%) & 252k & 91.53 & 88.64 \\ 
    PH-Multi-Scopic & 4 & 113k & (-74\%) & 141k & 91.59 & 89.65 \\ 
    PH-Multi-Scopic & 6 & 80k & (-81\%) & 106k & \textbf{93.14} & 89.59 \\ 
    PH-Multi-Scopic & 12 & 71k & (-83\%) & 98k & 90.55 & \textbf{92.03} \\ 
    \midrule[0.5pt]
    ResNet & -- & 926k & & 1,096k & 91.02 & 90.20\\
    PH-ResNet & 2 & 478k & (-48\%) & 607k & 91.50 & 89.88 \\
    PH-ResNet & 4 & 253k & (-73\%) & 361k & 90.89 & 90.91 \\
    PH-ResNet & 6 & 180k & (-81\%) & 281k & \textbf{91.95} & \textbf{90.77} \\
    PH-ResNet & 12 & 122k & (-87\%) & 173k & 90.82 & 89.78 \\
    \midrule[0.5pt]
    DenseNet & -- & 499k & & 511k & 90.03 & 89.10 \\
    PH-DenseNet & 2 & 253k & (-49\%) & 262k & 92.24 & 89.85\\
    PH-DenseNet & 4 & 132k & (-74\%) & 139k & 90.56 & 87.90\\
    PH-DenseNet & 6 & 95k & (-81\%) & 102k & 91.16 & 89.32\\
    PH-DenseNet & 12 & 96k & (-80\%) & 104k & \textbf{92.68} & \textbf{90.96} \\
    \bottomrule
\end{tabular}
\endgroup
\label{tab:results_12lead}
\end{table}

\section{Conclusion and Outlook}
\label{sec:conclusion}
We proposed a lightweight PH neural network framework for AF detection from ECG signals based on PH operations for multiplication and convolution.
We demonstrated comparable performance to corresponding real-valued networks for different CNN backbones on two datasets, while significantly reducing the parameter size. 
The dimensionality \(n\) can be chosen flexibly based on the data or tuned for further compression.
In future work, PH architectures can be combined with other advanced model compression techniques, such as knowledge distillation~\cite{Hinton2015}, to further reduce the
\mbox{model size.}
Implementation on wearable devices is further limited by the need for low computational complexity and power consumption, which should also be investigated in the future.
PH layers can also benefit the automatic analysis of other data types from wearable devices, such as photoplethysmography (PPG).

\bibliographystyle{IEEEtran}
\bibliography{my_abrv,IEEEabrv,refs}

\begin{thebibliography}{10}
\providecommand{\url}[1]{#1}
\csname url@samestyle\endcsname
\providecommand{\newblock}{\relax}
\providecommand{\bibinfo}[2]{#2}
\providecommand{\BIBentrySTDinterwordspacing}{\spaceskip=0pt\relax}
\providecommand{\BIBentryALTinterwordstretchfactor}{4}
\providecommand{\BIBentryALTinterwordspacing}{\spaceskip=\fontdimen2\font plus
\BIBentryALTinterwordstretchfactor\fontdimen3\font minus
  \fontdimen4\font\relax}
\providecommand{\BIBforeignlanguage}[2]{{%
\expandafter\ifx\csname l@#1\endcsname\relax
\typeout{** WARNING: IEEEtran.bst: No hyphenation pattern has been}%
\typeout{** loaded for the language `#1'. Using the pattern for}%
\typeout{** the default language instead.}%
\else
\language=\csname l@#1\endcsname
\fi
#2}}
\providecommand{\BIBdecl}{\relax}
\BIBdecl

\bibitem{Lippi2021}
G.~Lippi, F.~Sanchis-Gomar, and G.~Cervellin, ``Global epidemiology of atrial
  fibrillation: An increasing epidemic and public health challenge,''
  \emph{Int. J. Stroke}, vol.~16, pp. 217--221, 2021.

\bibitem{Rizwan2021}
A.~Rizwan, A.~Zoha, I.~B. Mabrouk, H.~M. Sabbour, A.~S. Al-Sumaiti,
  A.~Alomainy, M.~A. Imran, and Q.~H. Abbasi, ``A review on the state of the
  art in atrial fibrillation detection enabled by machine learning,''
  \emph{{IEEE} Rev. Biomed. Eng.}, vol.~14, pp. 219--239, 2021.

\bibitem{sajeev2019wearable}
J.~K. Sajeev, A.~N. Koshy, and A.~W. Teh, ``Wearable devices for cardiac
  arrhythmia detection: A new contender?'' \emph{Intern. Med. J.}, vol.~49,
  no.~5, pp. 570--573, 2019.

\bibitem{Hong2020}
S.~Hong, Y.~Zhou, J.~Shang, C.~Xiao, and J.~Sun, ``Opportunities and challenges
  of deep learning methods for electrocardiogram data: A systematic review,''
  \emph{Comput. Biol. Med.}, vol. 122, p. 103801, 2020.

\bibitem{Hannun2019}
A.~Y. Hannun, P.~Rajpurkar, M.~Haghpanahi, G.~H. Tison, C.~Bourn, M.~P.
  Turakhia, and A.~Y. Ng, ``Cardiologist-level arrhythmia detection and
  classification in ambulatory electrocardiograms using a deep neural
  network,'' \emph{Nat. Med.}, vol.~25, pp. 65--69, 2019.

\bibitem{Andersen2019}
R.~S. Andersen, A.~Peimankar, and S.~Puthusserypady, ``A deep learning approach
  for real-time detection of atrial fibrillation,'' \emph{Expert Syst. Appl.},
  vol. 115, pp. 465--473, 2019.

\bibitem{Han2015}
S.~Han, H.~Mao, and W.~J. Dally, ``Deep compression: Compressing deep neural
  networks with pruning, trained quantization and {Huffman} coding,'' in
  \emph{Proc.\ ICLR}, 2015.

\bibitem{Hinton2015}
G.~Hinton, O.~Vinyals, and J.~Dean, ``Distilling the knowledge in a neural
  network,'' \emph{arXiv preprint arXiv:1503.02531}, 2015.

\bibitem{polino2018model}
A.~Polino, R.~Pascanu, and D.~Alistarh, ``Model compression via distillation
  and quantization,'' in \emph{Proc.\ ICLR}, 2018, pp. 1--21.

\bibitem{Parcollet2020}
T.~Parcollet, M.~Morchid, and G.~Linarès, ``A survey of quaternion neural
  networks,'' \emph{Artif. Intell. Rev.}, vol.~53, pp. 2957--2982, 2020.

\bibitem{Zhang2021}
A.~Zhang, Y.~Tay, S.~Zhang, A.~Chan, A.~T. Luu, S.~C. Hui, and J.~Fu, ``Beyond
  fully-connected layers with quaternions: Parameterization of hypercomplex
  multiplications with $1/n$ parameters,'' in \emph{Proc.\ ICLR}, 2021, pp.
  1--13.

\bibitem{Grassucci2021}
E.~Grassucci, A.~Zhang, and D.~Comminiello, ``{PHNNs}: Lightweight neural
  networks via parameterized hypercomplex convolutions,'' \emph{{IEEE} Trans.
  Neural Networks Learn. Syst.}, pp. 1--13, 2022.

\bibitem{cruces2018quaternion}
P.~D. Cruces, R.~Correa, E.~Laciar, and P.~Arini, ``{Quaternion neural network
  with temporal feedback calculation: Application to cardiac vector velocity
  during myocardial infarction},'' \emph{Revista Argentina de
  Bioingenier{\'\i}a}, vol.~22, no.~3, pp. 60--64, 2018.

\bibitem{Hu2018}
J.~Hu, L.~Shen, and G.~Sun, ``Squeeze-and-excitation networks,'' in
  \emph{Proc.\ CVPR}, 2018, pp. 7132--7141.

\bibitem{de2020temporal}
R.~R. De~With, {\"O}.~Erk{\"u}ner, M.~Rienstra, B.-O. Nguyen, F.~W. K{\"o}rver,
  D.~Linz, H.~Cate~Ten, H.~Spronk, A.~A. Kroon, A.~H. Maass \emph{et~al.},
  ``{Temporal patterns and short-term progression of paroxysmal atrial
  fibrillation: Data from RACE V},'' \emph{EP Europace}, vol.~22, no.~8, pp.
  1162--1172, 2020.

\bibitem{Cai2020}
W.~Cai and D.~Hu, ``{QRS} complex detection using novel deep learning neural
  networks,'' \emph{IEEE Access}, vol.~8, pp. 97\,082--97\,089, 2020.

\bibitem{He2016}
K.~He, X.~Zhang, S.~Ren, and J.~Sun, ``Deep residual learning for image
  recognition,'' in \emph{Proc.\ CVPR}, 2016, pp. 770--778.

\bibitem{Huang2017}
G.~Huang, Z.~Liu, L.~van~der Maaten, and K.~Q. Weinberger, ``Densely connected
  convolutional networks,'' in \emph{Proc.\ CVPR}, 2017, pp. 4700--4708.

\bibitem{torch_ecg_paper2}
H.~Wen and J.~Kang, ``A novel deep learning package for electrocardiography
  research,'' \emph{Physiol. Meas.}, vol.~43, no.~11, p. 115006, 2022.

\bibitem{Wang2021}
\BIBentryALTinterwordspacing
X.~Wang, C.~Ma, X.~Zhang, H.~Gao, G.~D. Clifford, and C.~Liu, ``Paroxysmal
  atrial fibrillation events detection from dynamic {ECG} recordings: The 4th
  {China Physiological Signal Challenge} 2021,'' 2021. [Online]. Available:
  \url{http://2021.icbeb.org/CPSC2021}
\BIBentrySTDinterwordspacing

\bibitem{Wen2022:cpsc2021-work}
H.~Wen and J.~Kang, ``A comparative study on neural networks for paroxysmal
  atrial fibrillation events detection from electrocardiography,'' \emph{J.
  Electrocardiol.}, vol.~75, pp. 19--27, 2022.

\bibitem{CPSC2018}
F.~Liu, C.~Liu, L.~Zhao, X.~Zhang, X.~Wu, X.~Xu, Y.~Liu, C.~Ma, S.~Wei, Z.~He
  \emph{et~al.}, ``An open access database for evaluating the algorithms of
  electrocardiogram rhythm and morphology abnormality detection,'' \emph{J.
  Med. Imaging Health Inf.}, vol.~8, pp. 1368--1373, 2018.

\bibitem{CinC2020}
\BIBentryALTinterwordspacing
E.~A.~P. Alday, A.~Gu, A.~J. Shah, C.~Robichaux, A.~K.~I. Wong, C.~Liu, F.~Liu,
  A.~B. Rad, A.~Elola, S.~Seyedi \emph{et~al.}, ``Classification of 12-lead
  {ECGs}: the {PhysioNet/Computing in Cardiology Challenge} 2020,''
  \emph{Physiol. Meas.}, vol.~41, p. 124003, 2020. [Online]. Available:
  \url{https://moody-challenge.physionet.org/2020/}
\BIBentrySTDinterwordspacing

\bibitem{PhysioNet}
A.~L. Goldberger, L.~A.~N. Amaral, L.~Glass, J.~M. Hausdorff, P.~C. Ivanov,
  R.~G. Mark, J.~E. Mietus, G.~B. Moody, C.-K. Peng, and H.~E. Stanley,
  ``{PhysioBank, PhysioToolkit, and PhysioNet}: Components of a new research
  resource for complex physiologic signals,'' \emph{Circulation}, vol. 101,
  no.~23, pp. e215--e220, 2000.

\bibitem{Ribeiro2020:ecg}
A.~H. Ribeiro, M.~H. Ribeiro, G.~M.~M. Paix{\~a}o, D.~M. Oliveira, P.~R. Gomes,
  J.~A. Canazart, M.~P.~S. Ferreira, C.~R. Andersson, P.~W. Macfarlane,
  W.~Meira~Jr. \emph{et~al.}, ``Automatic diagnosis of the 12-lead {ECG} using
  a deep neural network,'' \emph{Nat. Commun.}, vol.~11, no.~1, p. 1760, 2020.

\bibitem{torch_ecg}
\BIBentryALTinterwordspacing
H.~Wen and J.~Kang, ``torch\_ecg: An {ECG} deep learning framework implemented
  using {PyTorch},'' 2022, v0.0.26. [Online]. Available:
  \url{https://github.com/DeepPSP/torch_ecg}
\BIBentrySTDinterwordspacing

\bibitem{Ridnik2021_AsymmetricLoss}
T.~Ridnik, E.~Ben-Baruch, N.~Zamir, A.~Noy, I.~Friedman, M.~Protter, and
  L.~Zelnik-Manor, ``Asymmetric loss for multi-label classification,'' in
  \emph{Proc.\ ICCV}, 2021, pp. 82--91.

\bibitem{Reddi2018_amsgrad}
S.~J. Reddi, S.~Kale, and S.~Kumar, ``On the convergence of adam and beyond,''
  in \emph{Proc.\ ICLR}, 2018, pp. 1--23.

\end{thebibliography}

\end{document}